\documentstyle[12pt,epsf]{article}
\begin{document}

\baselineskip 18pt
\setcounter{page}{0}

\newpage

\pagenumbering{arabic}
\setcounter{page}{1}
\baselineskip 12pt

\leftskip 10cm
{\bf
DPNU-94-32\\
KEK Preprint 94-101\\
TIT-HPE-94-07\\
TUAT-HEP 94-3\\
INS-REP-1062\\
NWU-HEP 94-03\\
OCU-HEP 94-5\\
KOBE-HEP 94-04\\
PU-94-684
}

\leftskip 0cm
\baselineskip 21pt

\renewcommand{\thefootnote}{\fnsymbol{footnote}}

{\flushleft
{\Large \bf Measurement of the forward-backward asymmetries for charm-
and bottom-quark pair productions at $\langle \sqrt{s} \rangle$=58GeV with
electron tagging
\footnote{to be published in Physics Letters B
}
}
}
\vskip 1.5cm

{
TOPAZ Collaboration\\
\underline {E.Nakano$^{a,}$}
\footnote{internet address;NAKANOE@KEKVAX.KEK.JP}
R.Enomoto$^b$, M.Iwasaki$^c$,
K.Abe$^a$, T.Abe$^a$, I.Adachi$^b$,
K.Adachi$^c$, M.Aoki$^d$, M.Aoki$^a$, S.Awa$^c$, R.Belusevic$^b$,
K.Emi$^e$, H.Fujii$^b$, K.Fujii$^b$,T.Fujii$^f$, J.Fujimoto$^b$,
 K.Fujita$^g$, N.Fujiwara$^c$, H.Hayashii$^c$,
B.Howell$^h$, N.Iida$^b$, H.Ikeda$^b$, R.Itoh$^b$, Y.Inoue$^g$, H.Iwasaki$^b$,
K.Kaneyuki$^d$, R.Kajikawa$^a$,
S.Kato$^i$, S.Kawabata$^b$, H.Kichimi$^b$, M.Kobayashi$^b$,
 D.Koltick$^h$, I.Levine$^h$, S.Minami$^d$,
K.Miyabayashi$^a$, A.Miyamoto$^b$, K.Muramatsu$^c$, K.Nagai$^j$,
K.Nakabayashi$^a$, O.Nitoh$^e$, S.Noguchi$^c$, A.Ochi$^d$, F.Ochiai$^k$,
 N.Ohishi$^a$, Y.Ohnishi$^a$, Y.Ohshima$^d$,
H.Okuno$^i$, T.Okusawa$^g$,T.Shinohara$^e$, A.Sugiyama$^a$,
S.Suzuki$^a$, S.Suzuki$^d$, K.Takahashi$^e$, T.Takahashi$^g$,
 T.Tanimori$^d$, T.Tauchi$^b$, Y.Teramoto$^g$, N.Toomi$^c$,
 T.Tsukamoto$^b$, O.Tsumura$^e$, S.Uno$^b$, T.Watanabe$^d$,
Y.Watanabe$^d$, A.Yamaguchi$^c$, A.Yamamoto$^b$,and M.Yamauchi$^b$\\
}

\newpage

{\small \it
\leftline{(a)
    Department of Physics, Nagoya University, Nagoya 464, Japan}
\leftline{(b)
    KEK, National Laboratory for High Energy Physics, Ibaraki-ken 305, Japan }
\leftline{(c)
    Department of Physics, Nara Women's University, Nara 630, Japan }
\leftline{(d)
    Department of Physics, Tokyo Institute of Technology, Tokyo 152, Japan}
\leftline{(e)
    Department of Applied Physics, Tokyo Univ. of Agriculture and
    Technology, Tokyo 184, Japan}
\leftline{(f)
    Department of Physics, University of Tokyo, Tokyo 113, Japan}
\leftline{(g)
    Department of Physics, Osaka City University, Osaka 558, Japan }
\leftline{(h)
    Department of Physics, Purdue University, West Lafayette, IN 47907, USA }
\leftline{(i)
    Institute for Nuclear Study, University of Tokyo, Tanashi,
     Tokyo 188, Japan }
\leftline{(j)
    The Graduate School of Science and Technology, Kobe University, Kobe 657,
    Japan }
\leftline{(k)
    Faculty of Liberal Arts, Tezukayama University, Nara 631, Japan }
}

\vskip 2cm

\begin{abstract}
We have measured, with electron tagging, the forward-backward
asymmetries of charm- and
bottom-quark pair productions
at $\langle \sqrt{s} \rangle$=58.01GeV, based on
23,783 hadronic events selected from a data sample
of 197pb$^{-1}$ taken with the TOPAZ detector at TRISTAN.
The measured forward-backward asymmetries
are
$A_{FB}^c = -0.49 \pm 0.20(stat.) \pm 0.08 (sys.)$ and
$A_{FB}^b = -0.64 \pm 0.35(stat.) \pm 0.13 (sys.)$, which are
consistent with the standard model predictions.

\end{abstract}

\newpage

\section{Introduction}
\hspace*{\parindent}
The differential cross section for fermion pair productions, $e^+e^-
\rightarrow f \bar f$, can be written in the following form in the massless
limit:
$$
\frac{d\sigma_{f\bar f}}{d\cos \theta} =
\frac{3}{8}\sigma_{f\bar f}(1+\cos^2 \theta +
\frac{8}{3}A_{FB}^f\cos \theta),
$$
where $\sigma_{f\bar f}$ and $A_{FB}^f$ are
the total cross section and the forward-backward charge
asymmetry, respectively, while $\theta$ is the polar-angle
of the final-state fermion $f$ with respect to the direction of
the initial-state electron.

In the standard model\cite{sm}, $A_{FB}^f$ is given by
$$
A_{FB}^f = \frac{3}{4}\cdot
\frac{-2Q_fa_ea_f\Re(\chi)+4a_ev_ea_fv_f|\chi|^2}
{Q_f^2-2Q_fv_ev_f\Re(\chi)+(v_e^2+a_e^2)(v_f^2+a_f^2)|\chi|^2}
$$
with
$$
\chi =\frac
{s}
{16\sin^2\theta_W\cos^2\theta_W(s-M_{Z^0}^2+iM_{Z^0}\Gamma_{Z^0}^0)},
$$
where $v_e(v_f)$ and $a_e(a_f$) are the vector and axial-vector coupling
constants of the electron (final-state fermion) to the $Z^0$ boson, $Q_f$ is
the charge of the final-state fermion, and $M_{Z^0}$ and $\Gamma_{Z^0}^0$ are
the mass
and the total width of the $Z^0$ boson, respectively.
The standard model predicts
$$
v_f = 2(I_3^f - 2Q_f\sin^2\theta_W)
$$
and
$$
a_f = 2I_3^f,
$$
where $I_3^f$ is the third component of the weak isospin of
the fermion ($f$) and $\theta_W$ is the Weinberg angle.

The above formula tells us that $A_{FB}^f$ attains to its maximum
in the TRISTAN energy region
and that its measurement there is sensitive to $a_f$ and
therefore to the structure of the multiplet to which the fermion belongs.
The measurement of the forward-backward charge asymmetry thus provides
a good test of the standard model.
The predicted asymmetries for charm- and bottom-quark pair productions are
\begin{eqnarray}
\nonumber A_{FB}^c & = & -0.47,\\
\nonumber A_{FB}^b & = & -0.59
\end{eqnarray}
at $\sqrt s$=58.01GeV for
$M_{Z^0}=91.1888$GeV/c$^2$, $\Gamma_{Z^0}=2.4974$GeV,
and $\sin^2\theta_W=0.2321$
\cite{glasgow}.
The $A_{FB}^b$ is, however,
reduced by the $B$-$\bar B$ mixing, whose
probability $\chi$ is given by
$\chi = R_d\chi_d + R_s\chi_s$, where $R_i$, and $\chi_i$ are the fraction
and the mixing probability of $B_d$ or $B_s$, respectively.
Their measured values\cite{pdg}
are
$$
\chi_d = 0.16, R_d = \frac{1.0}{2.3}
$$
and
$$
\chi_s = 0.53, R_s = \frac{0.3}{2.3},
$$
so that the expected $A_{FB}^b$ value becomes $-0.43$.

In this energy region, only the data from the TRISTAN experiments
are available and
the TRISTAN average values of $A_{FB}^c$\cite{dstar,evenus,dvenus}
and $A_{FB}^b$\cite{mamy,shimonaka,Nagai,evenus,emvenus} have been
\begin{eqnarray}
\nonumber A_{FB}^c & = & -0.56 \pm 0.09,\\
\nonumber A_{FB}^b & = & -0.59 \pm 0.09.
\end{eqnarray}
The $A_{FB}^c$ is consistent with the standard model predictions
and the $A_{FB}^b$ is slightly deviated from it.

The TRISTAN data include our previous measurement\cite{dstar} of
the forward-backward charge asymmetry for the charm-quark pair
production through
both exclusive and inclusive
reconstructions of $D^{*\pm} \rightarrow \pi^\pm D^0$:
$A_{FB}^c = -0.49^{+0.14}_{-0.13}(stat.)\pm0.06(sys.)$,
consistent with the standard model
prediction.

In order to improve the statistical accuracy,
we have carried out another measurements using
electron tagging
to be described in this paper.
It should be noted that this measurement is completely independent of
the above $D^*$ analysis and that, in our energy region, there is only one
previous $A_{FB}^c$ measurement reported, which are used
lepton tagging at this $\sqrt s$
\cite{evenus}.

\section{The TOPAZ detector}
\hspace*{\parindent}
The main components of the TOPAZ detector\cite{topaz}
include a time projection chamber (TPC)
and a barrel lead-glass calorimeter (BCL)
which were essential to electron identification.

Combining the dE/dx information from the TPC and the E/P information
from the BCL, we were able to select electrons in hadronic final states
with high purity and
high efficiency over a broad momentum range.

Instead of getting into details of these detectors, we summarize
their performance here.
The momentum resolution of the TPC has been measured to be
$$
\frac{\sigma_{P_{t}}}{P_{t}} =
\sqrt{1.0+1.0P_{t}^2(GeV/c)^2}~~\%
$$
through $e^+e^- \rightarrow \mu^+\mu^-$ and cosmic $\mu^\pm$
events\cite{topaz},
while its dE/dx resolution was determined to be 4.6\% by a study of
minimum ionizing pions.
The energy resolution of the BCL can, on the other hand, be expressed as
$$
\frac{\sigma_E}{E} =
\sqrt{\biggl({8.0\over\sqrt{E(GeV)}}\biggr)^2+(1.5)^2}~~\%.
$$

\section{Analysis}
\subsection{Electron selection}
\hspace*{\parindent}
This analysis is based on 23,783 hadronic events.
The selection method was described in
Ref\cite{hadsel}.
This data sample corresponded to an integrated
luminosity of 197pb$^{-1}$ and was taken at an averaged center-of-mass energy
$\langle \sqrt s \rangle$=58.01GeV.

In search of electron track candidates, we first selected good charged tracks
from the hadronic events, using the following selection criteria defining
a good track.
\begin{enumerate}
\item The closest approach to the interaction point ($R$) had to be
less than 1.0cm
in the X-Y plane (perpendicular to the beam axis) and that
in the Z direction ($Z$) to be less than 4.0cm.
\item the absolute value of the cosine of the polar angle had to be between
0.02 and 0.83,
\item the transverse momentum ($P_t$) with respect to the beam axis had to be
greater than 0.15GeV/c,
\item the number of degrees of freedom (N.D.F.) in the track fitting had to be
greater than 3, and
\item there had to be more than 30 hit wires for dE/dx calculation
(65\%-truncated mean)
out of 114 wires maximum.
\end{enumerate}

Each good track was extrapolated to the BCL to look for its
corresponding BCL cluster and to test E/P.
The clustering of the energy deposits in the BCL was carried out
iteratively by merging
a counter to its neighboring counter if its energy was
smaller than that of the neighboring counter.
For each of so formed clusters, we calculated its energy
as the sum over counters
and its position as the energy-weighted mean of
the counter positions.

\begin{figure}
\vskip -1.5cm
\epsfysize9.1cm
\hskip 1in
\epsfbox{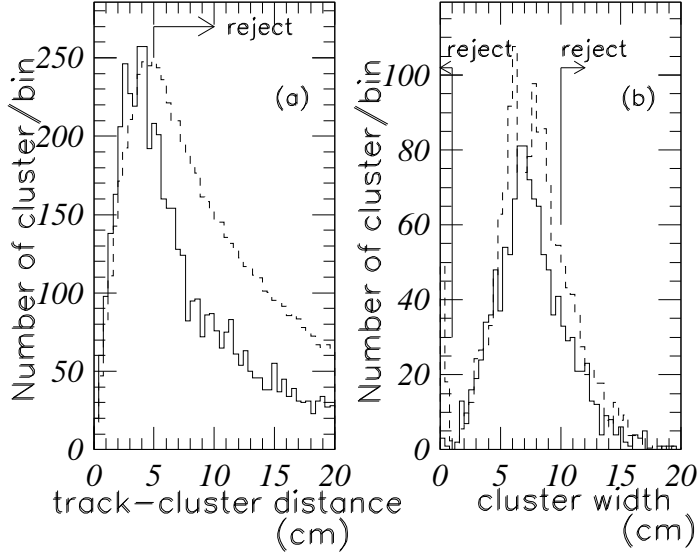}
\vskip -0.6cm
\caption{\footnotesize
(a) The distribution of the minimum distance between the extrapolated track
position on the BCL and the cluster position.
(b) The distribution of the cluster width of a shower with respect to
the track-extrapolated position.
The solid and dashed histograms are for the electron-enhanced sample defined
in the text and for all tracks, respectively.
}
\label{cls}
\end{figure}

Comparing the extrapolated track position and the BCL cluster positions,
we looked for the cluster that is the closest to the track.
The closest distances between tracks and clusters are
histogrammed in Fig.\ref{cls}-(a)
for all tracks (dashed) and for an electron-enhanced sample (solid)
selected by requiring
0.75$<$E/P$<$1.25 and 5.5$<$dE/dx$<$7.5(keV/cm).
We accepted those tracks which had a distance less than 5.0 cm.

We also calculated the energy-weighted r.m.s. of counter positions in
the cluster (cluster width) with respect to the matched track position,
whose distribution is shown in Fig.\ref{cls}-(b).
The selected tracks were further required to have
a cluster width between 1.0 and 10.0cm.

Since background tracks were predominantly in
the low-momentum region, we imposed an additional momentum cut
$P>0.8$GeV/c.

Finally, the dE/dx information from the TPC was used to complete our
electron selection:
the $\chi^2$ for electron hypothesis ($\chi^2_e$) had
to be $\chi^2_e < 3.0$ (N.D.F. = 1). The tracks rejected by this cut
(dE/dx-rejected hadrons) are to be used in the background estimation.

\subsection{Rejection of $e^+e^-$ pair background}
\hspace*{\parindent}
We rejected electrons apparently coming from $\gamma$ conversions or
Dalitz decays as follows.
Secondary vertices were reconstructed for all unlike-sign pairs of the TPC
tracks.
When the two tracks in a pair did not intersect in the X-Y plane,
we required the pair to have
a distance at the closest point less than 4.0cm in the X-Y plane, and
2.0cm in the Z direction.
The pair also had
to have a deflection angle in the X-Y plane less than 5.0 degrees
between its momentum vector and the flight direction
from the interaction point.
If the invariant mass of the pair was less than 50MeV/c$^2$,
the tracks were rejected.

When the tracks intersected in the X-Y plane,
we selected from the two intersections
the one which gave the smaller deflection angle as the secondary
vertex.
We then required the pair to have a distance less than 2.0cm
in the Z direction and to have a deflection angle
less than 5.0 degrees.
If the invariant mass was less than 200MeV/c$^2$,
the tracks were rejected.

In this way, most of the electrons coming from $\gamma$ conversions or
Dalitz decays were removed.
Nevertheless, there still remained a significant number of electrons
from $\gamma$ conversions
or Dalitz decays, which were
estimated through Monte-Carlo simulation\cite{lund}:
the Monte-Carlo simulation gave us
the ratio of
the number of all the reconstructed conversion tracks to that of
the remaining tracks.

Using this ratio, we estimated the number of remaining
pair-conversion tracks from
the actual number of all the reconstructed pairs in the
experimental data
on a bin by bin basis and subtracted them from
the electron candidates.
By doing this, we can reduce the systematic errors due to the error of
the material thickness in the detector simulation program.
The remaining electrons from $\gamma$ conversions or Dalitz decays
estimated through the Monte-Carlo simulation are $311.7 \pm 17.7$ events.
The E/P distribution for these electron candidates is shown in Fig.\ref{eop}.

\subsection{Hadron background}
\begin{figure}
\vskip-1.5cm
\epsfysize9.1cm
\hskip 1in
\epsfbox{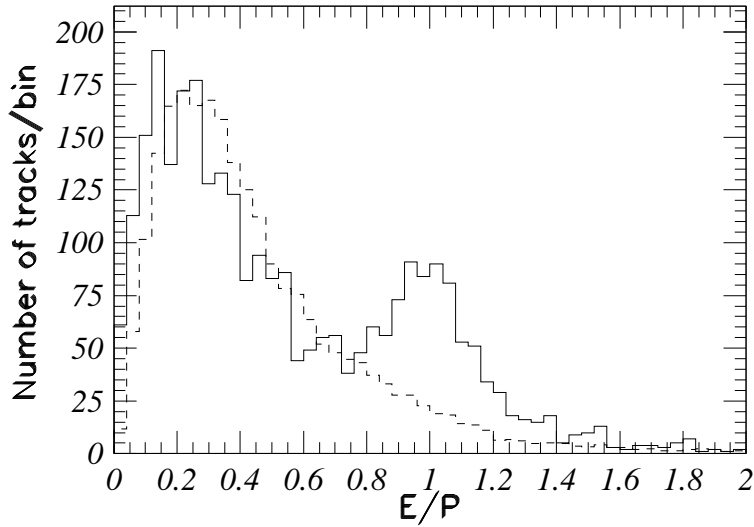}
\vskip -0.6cm
\caption{\footnotesize
The E/P distribution of the selected electron samples.
The solid histogram shows the experimental data. The dashed histogram is
for the estimated
background by dE/dx-rejected hadrons.
}
\label{eop}
\end{figure}

\hspace*{\parindent}
The hadron background was estimated using the dE/dx-rejected hadrons in
the experimental data
and is shown in Fig.\ref{eop} as the dashed histogram, whose
normalization factor was calculated so as to equalize
the entries in the side-band (E/P = 0.0 - 0.64) for
the electron candidates and the background sample.
The normalized background was subtracted from the electron candidates
and the remaining electrons in the region 0.72$<$E/P$<$2.00 were counted.
This method had been checked out through the Monte-Carlo simulation.

The estimated number of electrons
from primary charm, b-to-c cascade, and direct bottom decays
are $151.5 \pm 12.3$, $66.5 \pm 8.2$, and $131.9 \pm 11.5$ events,
where the selection efficiencies of 10.4\%, 11.3\%, and 24.1\%,
respectively.

\subsection{Charm- and bottom-quark sample}
\hspace*{\parindent}
We calculated the transverse momentum ($P_T$) of each electron candidate
track with respect
to the axis of jets reconstructed using the invariant-mass
algorithm \cite{dstar}.
For the jet reconstruction, we used charged tracks with
momenta greater than
0.2GeV/c and all of the neutral clusters, which were
clusters with a distance to the closest track greater than 5.0 cm.

We studied the angular resolution of the jet axis with respect to
the primary quark direction through the Monte-Carlo simulation.
Since the resolution of the thrust axis was about 9 degrees,
while that of the reconstructed jets as above was about
6 degrees, we used the reconstructed jets as the primary quark directions.

The $P_T$ of an electron from a charm-quark is expected to be lower than
that of a bottom-quark in general.
To enhance charm or bottom contents, therefore, we
divided the sample at $P_T = 0.8$GeV/c into two classes:
low-$P_T$ (charm-enhanced) and high-$P_T$ (bottom-enhanced) samples.

\subsection{Monte Carlo simulation}
\hspace*{\parindent}
We generated Monte-Carlo events using the JETSET6.3 generator
\cite{lund}
with $M_{Z^0} = 91.173$GeV/c$^2, \Gamma_{Z^0} = 2.487$GeV, and
$\sin ^2 \theta_W = 0.2325$\cite{pdg}.
For light quark events,
its parameters were tuned by a multi-parameter fit
of hadronic event shapes\cite{hadsel}.
For heavy quark events, we adjusted the parameters for fragmentation function,
so as to match other experiments\cite{evenus}, to be a=0.8 and b=0.2.
The $B$-$\bar B$ mixing effect is included in our Monte-Carlo simulation.
Using this Monte-Carlo simulation, we estimated the acceptance and
radiative correction factors to be used later in the following subsection.

\subsection{Fitting procedure}
\begin{figure}
\vskip -1.5cm
\epsfysize9.1cm
\hskip 1in
\epsfbox{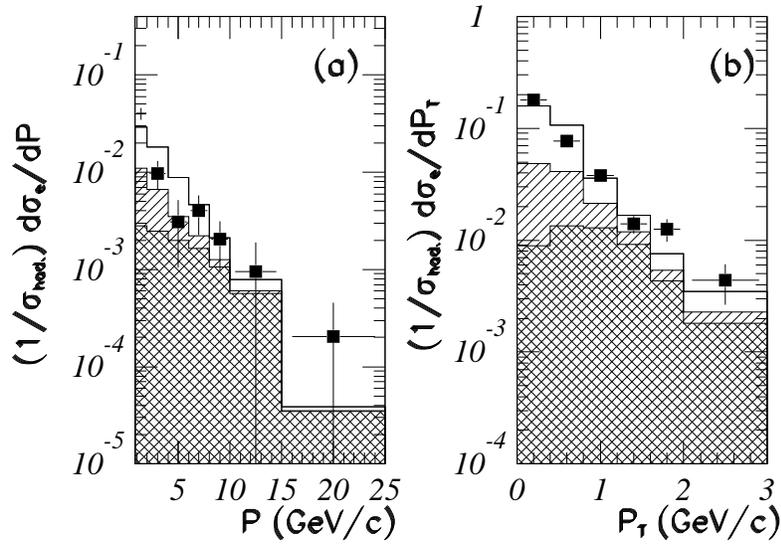}
\vskip -0.6cm
\caption{\footnotesize
(a) The momentum distribution of $e^\pm$'s.
(b) The $P_T$ distribution of $e^\pm$'s.
The points are background-subtracted data. The open, hatched,
and double-hatched histograms are the best-fit results for
the contributions from
$c \rightarrow e^+X$,
$b \rightarrow c \rightarrow e^+X$, and $b \rightarrow e^-X$, respectively.
}
\label{ppt}
\end{figure}

\begin{figure}
\vskip -1.5cm
\epsfysize9.1cm
\hskip 1in
\epsfbox{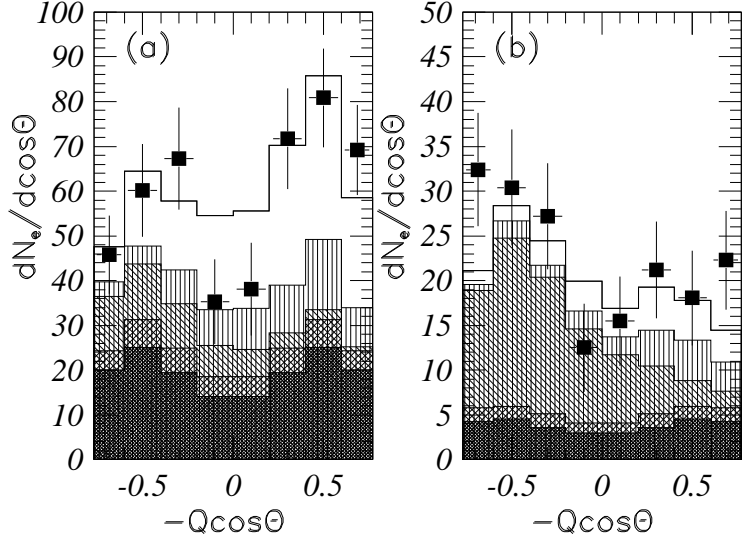}
\vskip -0.6cm
\caption{\footnotesize
The -Q$\cos\theta$ distributions of $e^\pm$'s (a)in the low-$P_T$ region
($P_T<0.8$GeV/c) and
(b)in the high-$P_T$ region
($P_T>0.8$GeV/c).
The points are experimental data.
The open, hatched, double-hatched, triple-hatched, and filled histograms are
the best-fit results for the contributions from
$c \rightarrow e^-X, \bar b \rightarrow \bar c \rightarrow e^-X$,
$ \bar c \rightarrow e^-X$, $\gamma$ conversion,
and Dalitz decay, respectively.
}
\label{cos}
\end{figure}

\hspace*{\parindent}
Figs.\ref{ppt}-(a) and -(b) show the momentum and the $P_T$ distributions of
electrons, respectively.
The $-Q\cos\theta$ distributions of the high-$P_T$ and the low-$P_T$ electrons
are shown in Figs.\ref{cos}-(a) and -(b), respectively,
where $Q$ is the charge of the electron and $\theta$ is the polar angle of
the jet axis with respect to the electron beam axis.
The points with error bars in the figures are experimental data, while
the histograms are the best-fit results obtained
by fitting Fig.\ref{ppt}-(a), Figs.\ref{cos}-(a), and -(b) simultaneously.
The used fit function to the $-Q\cos\theta$ distribution is
\begin{eqnarray}
\nonumber N^{e^\pm}_i &=&
          N^{exp}_{q \bar q}\frac{\sigma_{c \bar c}}{\sigma_{q \bar q}}\cdot
          2Br(c \rightarrow e)F(-A_{FB}^c)_i\cdot C^{pc}_i\\
\nonumber&+&N^{exp}_{q \bar q}\frac{\sigma_{b \bar b}}{\sigma_{q \bar q}}\cdot
          2Br(b \rightarrow e)F(A_{FB}^b)_i\cdot C^{pb}_i\\
\nonumber&+&N^{exp}_{q \bar q}\frac{\sigma_{b \bar b}}{\sigma_{q \bar q}}\cdot
          2Br(c \rightarrow e)(1+x)F(-\frac{1-x}{1+x}A_{FB}^b)_i\cdot C^{ca}_i
\end{eqnarray}
with $x \equiv Br(b \rightarrow c \bar c s)$ set to be 16\%\cite{ctos},
where, $N^{exp}_{q \bar q}$ is the number of hadronic events, and
$F(A_{FB}^j)_i$ are integrals of the following formula over $i$-th
$\cos \theta$ bin;
$$\frac{3}{8}(1+\cos^2\theta+\frac{8}{3}A_{FB}^j\cos\theta).$$
$C^{pc}_i,C^{pb}_i$, and $C^{ca}_i$ are the Monte-Carlo-determined
correction factors for the $i$-th bin
of electrons from
prompt charm, prompt bottom, and b-to-c cascade decays,
respectively, which were described in the previous subsection and they
are listed in Table \ref{corr}.
\begin{table}
\begin{center}
\begin{tabular}{cccccccc}\hline\hline
\multicolumn{2}{c}{$\cos \theta$ region} & \multicolumn{2}{c}{prompt charm}&
                      \multicolumn{2}{c}{prompt bottom}&
                      \multicolumn{2}{c}{b-to-c cascade}\\
           &          & \multicolumn{2}{c}{$C^{pc}_i$}&
			\multicolumn{2}{c}{$C^{pb}_i$}&
			\multicolumn{2}{c}{$C^{ca}_i$} \\
           &          & low-$P_T$ & high-$P_T$ &
                      low-$P_T$ & high-$P_T$ &
                      low-$P_T$ & high-$P_T$ \\
\hline
 -0.77& -0.6          & 0.0967& 0.0185& 0.1134& 0.1220& 0.1006& 0.0216\\
 -0.6 & -0.4          & 0.1640& 0.0164& 0.1156& 0.1748& 0.0971& 0.0470\\
 -0.4 & -0.2          & 0.1319& 0.0234& 0.1275& 0.1949& 0.1552& 0.0274\\
 -0.2 &  0.0          & 0.1430& 0.0224& 0.1182& 0.1785& 0.1358& 0.0340\\
  0.0 &  0.2          & 0.1212& 0.0177& 0.1398& 0.1761& 0.1147& 0.0247\\
  0.2 &  0.4          & 0.1373& 0.0211& 0.1196& 0.1865& 0.1063& 0.0399\\
  0.4 &  0.6          & 0.1308& 0.0160& 0.1096& 0.1437& 0.1258& 0.0362\\
  0.6 &  0.77         & 0.0830& 0.0120& 0.0682& 0.1607& 0.0696& 0.0253\\
\hline\hline
\end{tabular}
\caption{\footnotesize List of correction factors.}
\label{corr}
\end{center}
\end{table}

The fit determined the branching fractions of $c \rightarrow e$ and
$b \rightarrow e$ to be
\begin{eqnarray}
\nonumber Br(c \rightarrow e) & = & 0.131 \pm 0.015 \\
\nonumber Br(b \rightarrow e) & = & 0.109 \pm 0.025
\end{eqnarray}
and $A_{FB}^c$ and $A_{FB}^b$ to be
\begin{eqnarray}
\nonumber A_{FB}^c & = & -0.49 \pm 0.20 \\
\nonumber A_{FB}^b & = & -0.64 \pm 0.35.
\end{eqnarray}

\subsection{Systematic errors}
\hspace*{\parindent}
We checked various systematic-error sources.
The estimated systematic errors are summarized in Table \ref{syserr}
and the varied parameter values used for the estimation
are listed in Table \ref{syschg}.
The dependence on the selection of good tracks was checked
by changing the cut values
on $R,Z,P_{t}$, and momentum (P).
The dE/dx dependence was tested by changing the $\chi^2_e$ cut.
The error caused by the shower shape parameters was estimated
by changing the distance cut and the cluster width cut.
We checked the effect of the background estimation by changing
the region for background
normalization and electron counting:
the first set used 0.0$<$E/P$<$0.64 and 0.72$<$E/P$<$2.00
for normalization
and counting, respectively,
while the second one used 0.40$<$E/P$<$0.60 and
1.6$<$E/P$<$2.00 for normalization and 0.72$<$E/P$<$1.28 for counting.
The systematic error due to the estimation of the $\gamma$ conversions
and Dalitz decays was checked by changing the cuts for the $\gamma$
conversion
and Dalitz decay rejection.
The QCD correction was estimated by comparing the JETSET 6.3 P.S. and
JETSET 6.3 $q \bar q$ options to be -0.02 and -0.01
for $A_{FB}^c$ and
$A_{FB}^b$, respectively. Since they were small,
we neglected them.
The errors from the $P_T$ cut is also given in Table \ref{syserr}.
The errors due to uncertainty of $B$-$\bar B$ mixing were also checked and
they were negligibly small($\sim$ 0.2\%).
The overall systematic errors were obtained by adding them in quadrature.

\begin{table}
\begin{center}
\begin{tabular}{lcccc}\hline\hline
error source& $\frac{\Delta Br(c \rightarrow e)}{Br(c \rightarrow e)}$&
$\frac{\Delta A_{FB}^c}{A_{FB}^c}$&
$\frac{\Delta Br(b \rightarrow e)}{Br(b \rightarrow e)}$&
$\frac{\Delta A_{FB}^b}{A_{FB}^b}$
 \\
\hline
track selection               & 7.4\%& 3.3\%& 6.4\%&17.7\%\\
dE/dx                         & 1.0\%& 7.5\%& 1.3\%& 3.2\%\\
cluster selection             & 5.7\%& 6.8\%&10.5\%& 0.6\%\\
background                    & 2.2\%& 8.2\%& 1.7\%& 6.7\%\\
$\gamma$ and Dalitz rejection & 1.5\%& 1.0\%& 1.6\%& 1.5\%\\
$P_T$ cut		      & 3.0\%& 8.3\%& 8.1\%& 7.2\%\\
\hline
total                         &10.2\%&15.8\%&14.9\%&20.6\%\\
\hline\hline
\end{tabular}
\caption{\footnotesize Summary of systematic errors.}
\label{syserr}
\end{center}
\end{table}

\begin{table}
\begin{center}
\begin{tabular}{llcc}\hline\hline
error source & parameter & nominal value & varied value \\
\hline
track selection   & $R$       & 1.0cm     & 1.2cm     \\
                  & $Z$       & 4.0cm     & 3.0cm     \\
                  & $P_t$     & 0.15GeV/c & 0.20GeV/c \\
                  & $P$       & 0.8GeV/c  & 1.0GeV/c  \\
dE/dx             & $\chi^2_e$ & 3.0       & 2.7       \\
cluster selection & distance  & 5.0cm     & 7.5cm     \\
                  & width     & 1.0 - 10.0cm & 2.0 - 12.5cm \\
$\gamma$ and Dalitz rejection & X-Y distance & 4.0cm & 5.0cm \\
                              & Z distance   & 2.0cm & 2.5cm \\
                              & deflection angle & 5 degrees & 7 degrees \\
$P_T$ cut         &           & 0.80GeV/c & 0.75GeV/c \\
$B$-$\bar B$ mixing& $\chi$     & 0.139     & 0.0 \\
\hline\hline
\end{tabular}
\caption{\footnotesize Summary of varied parameter values.}
\label{syschg}
\end{center}
\end{table}

\subsection{Results and discussion}
\begin{figure}
\vskip -1.5cm
\epsfysize9.1cm
\hskip 1in
\epsfbox{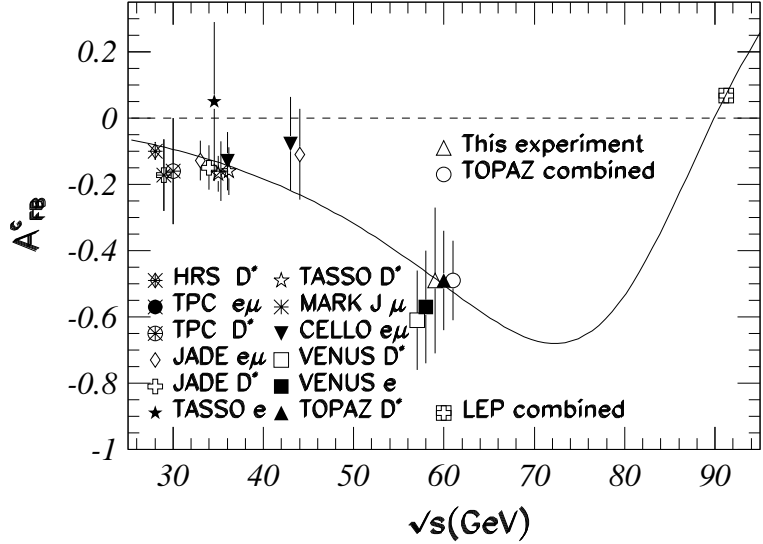}
\vskip -0.6cm
\caption{\footnotesize
Our $A_{FB}^c$ measurements together with other
experimental data shown as a function of \,\,~~~~
}
\vskip -0.5cm
\hskip 15.1cm
{\footnotesize$\sqrt{s}$}
\vskip 0.5cm

\label{acc}
\end{figure}

\hspace*{\parindent}
Our results are
\begin{eqnarray}
\nonumber Br ( c \rightarrow e ) = 0.131 \pm 0.015 (stat.) \pm 0.013 (sys.)\\
\nonumber Br ( b \rightarrow e ) = 0.109 \pm 0.025 (stat.) \pm 0.016 (sys.)\\
\nonumber A_{FB}^c = - 0.49 \pm 0.20 (stat.) \pm 0.08 (sys.)\\
\nonumber A_{FB}^b = - 0.64 \pm 0.35 (stat.) \pm 0.13 (sys.).
\end{eqnarray}
The branching fractions are consistent with the previous measurements, and
the forward-backward asymmetries are consistent with the standard model
predictions as well as our previous measurements
\cite{dstar,shimonaka,Nagai}.
The obtained forward-backward asymmetry for charm-quark pair production
is plotted in Fig. \ref{acc}
together with other experimental data
\cite{glasgow,dstar,evenus,dvenus,casym}.

The combined result with our previous $D^*$ analysis is
$$
A_{FB}^c = -0.49 \pm 0.12,
$$
from which we obtain the charm-quark's
axial-vector coupling constant($a^c_v$) to
the $Z^0$ boson
to be $a^c_v = 1.07^{+0.42}_{-0.31}$,
being consistent with the standard model.

\section{Conclusion}
\hspace*{\parindent}
We have measured the forward-backward asymmetries of charm- and
bottom-quark pair productions via $e^+e^-$ annihilations
through an inclusive electron
analysis at $\langle \sqrt{s} \rangle$=58.01GeV.
The number of hadronic events used for this analysis
is 23,783, corresponding to
an integrated luminosity of 197pb$^{-1}$.

The measured forward-backward asymmetries
are
$A_{FB}^c = -0.49 \pm 0.20(stat.) \pm 0.08 (sys.)$ and
$A_{FB}^b = -0.64 \pm 0.35(stat.) \pm 0.13 (sys.)$, consistent with the
standard model prediction and our previous measurements.
Combining the results from our previous $D^{*\pm}$ measurement,
we obtained $A_{FB}^c=-0.49 \pm 0.12$ and $a^c_v = 1.07^{+0.42}_{-0.31}$.
The obtained branching fractions are
$Br(c \rightarrow e) = 13.1 \pm 1.5(stat.) \pm 1.3(sys.)\%$ and
$Br(b \rightarrow e) = 10.9 \pm 2.5(stat.) \pm 1.6(sys.)\%$, which
are in good agreement with the previously measured values of
$9.6\pm0.9\%$\cite{argus}
and $10.8\pm0.5\%$\cite{argus,cleo}, respectively.

\section{Acknowledgement}
The authors thank Dr. M. Sakuda for discussions on the analysis.
They also thank the TRISTAN accelerator
staff for the successful operation of TRISTAN.
The authors appreciate
all of the engineers and technicians at KEK as well as the collaborating
institutions: H. Inoue, N. Kimura, K. Shiino, M. Tanaka, K. Tsukada, N. Ujiie,
and H. Yamaoka.

\end{document}